\documentclass{icrc2009}

\usepackage{graphicx}   
\usepackage{caption}    
\usepackage[font=footnotesize]{subfig} 
\usepackage{fixltx2e}
\usepackage{url}

\newcommand{\shorttitle}[1]%
{\markboth{Proceedings of the 31\MakeLowercase{$^{st}$} ICRC, {\L}\'{o}d\'{z} 2009}{#1} }
\newcommand{\etal}{\MakeLowercase{\textit{et al. }}} 

\begin{document}
\title{MAGIC-II Camera Slow Control Software}

\author{\IEEEauthorblockN{Burkhard Steinke\IEEEauthorrefmark{1},
			  Tobias Jogler\IEEEauthorrefmark{1} and Daniela Borla Tridon\IEEEauthorrefmark{1} for the MAGIC Collaboration
                            }\\
\IEEEauthorblockA{\IEEEauthorrefmark{1}Max-Planck-Insitut f\"{u}r Physik, D-80805 Munich, Germany}
}

\shorttitle{B. Steinke \etal MAGIC-II Camera Control}
\maketitle

\begin{abstract}
 The Imaging Atmospheric Cherenkov Telescope MAGIC I has recently been extended to a stereoscopic system by adding a second 17 m  telescope, MAGIC-II. One of the major improvements of the second telescope is an improved camera. The \textit{Camera Control Program} is embedded in the telescope control software as an independent subsystem.\\
The \textit{Camera Control Program} is an effective software to monitor and control the camera values and their settings and is written in the visual programming language \mbox{LabVIEW}. The two main parts, the \textit{Central Variables File}, which stores all information of the pixel and other camera parameters, and the \textit{Comm Control Routine}, which controls changes in possible settings, provide a reliable operation. A safety routine protects the camera from misuse by accidental commands, from bad weather conditions and from hardware errors by automatic reactions. 
  \end{abstract}

 
\section{Introduction}
The Imaging Atmospheric Cherenkov Telescope MAGIC I, located at the Canary Island of La Palma, has recently been extended to a stereoscopic system by adding a second 17 m  telescope, MAGIC-II.
The two telescope system compared to a single one, is designed to provide an improved sensitivity in the stereoscopic operation mode and lower the energy threshold. In this paper  the slow control architecture of the MAGIC-II camera is described. The focus is set on the structure of the \textit{Camera Control Program} and its main routines and procedures.
\\

\section{Magic-II Camera}
\begin{figure}[!t]
  \centering
  \includegraphics[width=3.1in]{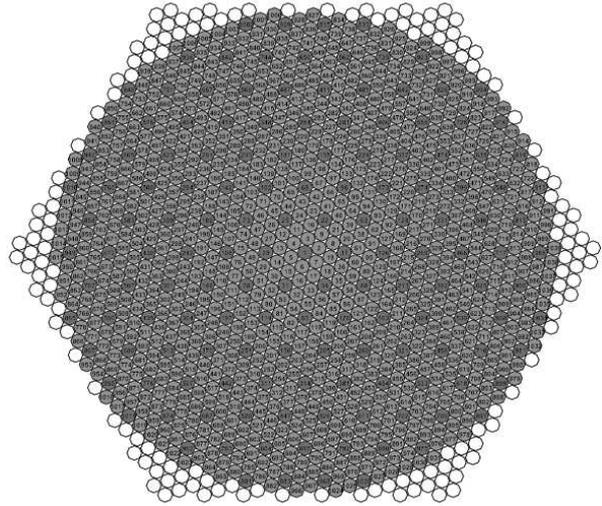}
  \caption{Scheme of the MAGIC-II camera PMT layout. The 1039 PMT pixel of $0.1^{o}$ diameter each form the circular, uniform camera.}
  \label{fig:camera_pixel_layout}
 \end{figure}

The camera of the telescope is placed in the focus of the parabolic reflector at a distance of 17.5 m from the elevation axis of the telescope structure.\\
The MAGIC-II camera is equipped uniformly with 1039 pixels of $0.1^{o}$ diameter each, covering a trigger radius of $1.25^{o}$ and a FoV of $3.5^{o}$ (cf. fig. \ref{fig:camera_pixel_layout}). Every seven pixels are grouped in a hexagonal configuration to form one cluster.
Each pixel consists of a 1" hemispherical, six dynode PMT, a Crockroft-Walton HV supply, an AC-coupled high bandwidth preamplifier to compensate the low gain of the PMT, a vertical-cavity surface-emitting laser (VCSEL) to convert the PMT signal into an optical signal to be transmitted by an optical fiber to the counting house, an anode current monitor, a pulse injector, temperature sensors and a monitor diode for the VCSEL light output. On the front side the pixels are equipped with Winston cone type light guides to minimize the dead area between the PMTs. The camera electronics is powered by two 5 V power supplies mounted in two boxes placed on the lower part of the camera, outside of the housing.
\section{The slow control architecture}
\begin{figure*}[th]
  \centering
  \includegraphics[width=4in]{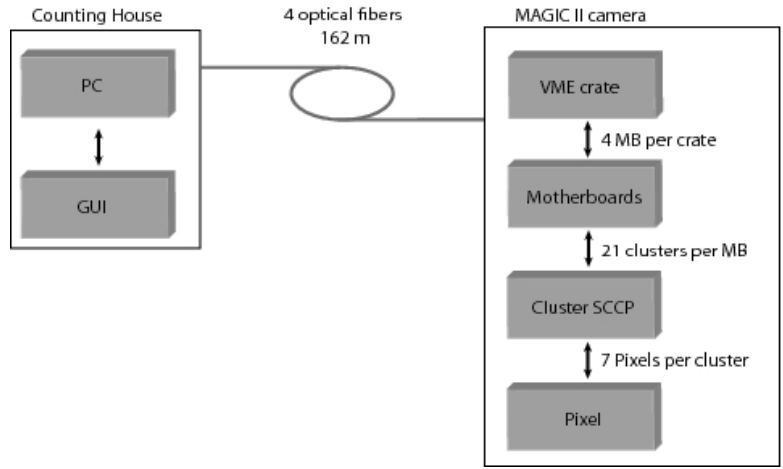}
  \caption{Scheme of the MAGIC-II camera control chain. 7 PMTs (pixels) are controlled by one Cluster SCCP. 22 Cluster SCCP are connected to a motherboard (MB), 4 motherboards at a time are adjunct to a VME interface crate. 4 optical fibres link the two used VME crates to the control PC.
    }
  \label{fig:control_chain}
 \end{figure*}

The camera is controlled by a slow control cluster processor (SCCP). 
A SCCP board installed in each cluster controls the operations of the camera and reads several parameters. The HV of each pixel can be set individually and the PMT current, the HV and the temperature at the VCSEL can be continuously monitored. A test-pulse generator board installed in each PMT to test the electrical chain is also controlled at this stage. In addition the slow control operates the lids in front of the Plexiglas window that protects the PMTs and steers the power supplies of the camera.\\
The SCCP has a flash programmable processor with 12 bits resolution DACs in a voltage range of 0 to 1.25 V and 12 bits resolution ADC in the range 0-2.5 V.
Each SCCP board is connected to a VME board in one of the two VME crates in the upper and lower part of the camera.\\
The VME crates are connected to the camera control PC in the counting house via an optical PCI to VME link (cf. fig. \ref{fig:control_chain}).

\begin{figure*}[th]
  \centering
  \includegraphics[width=5in]{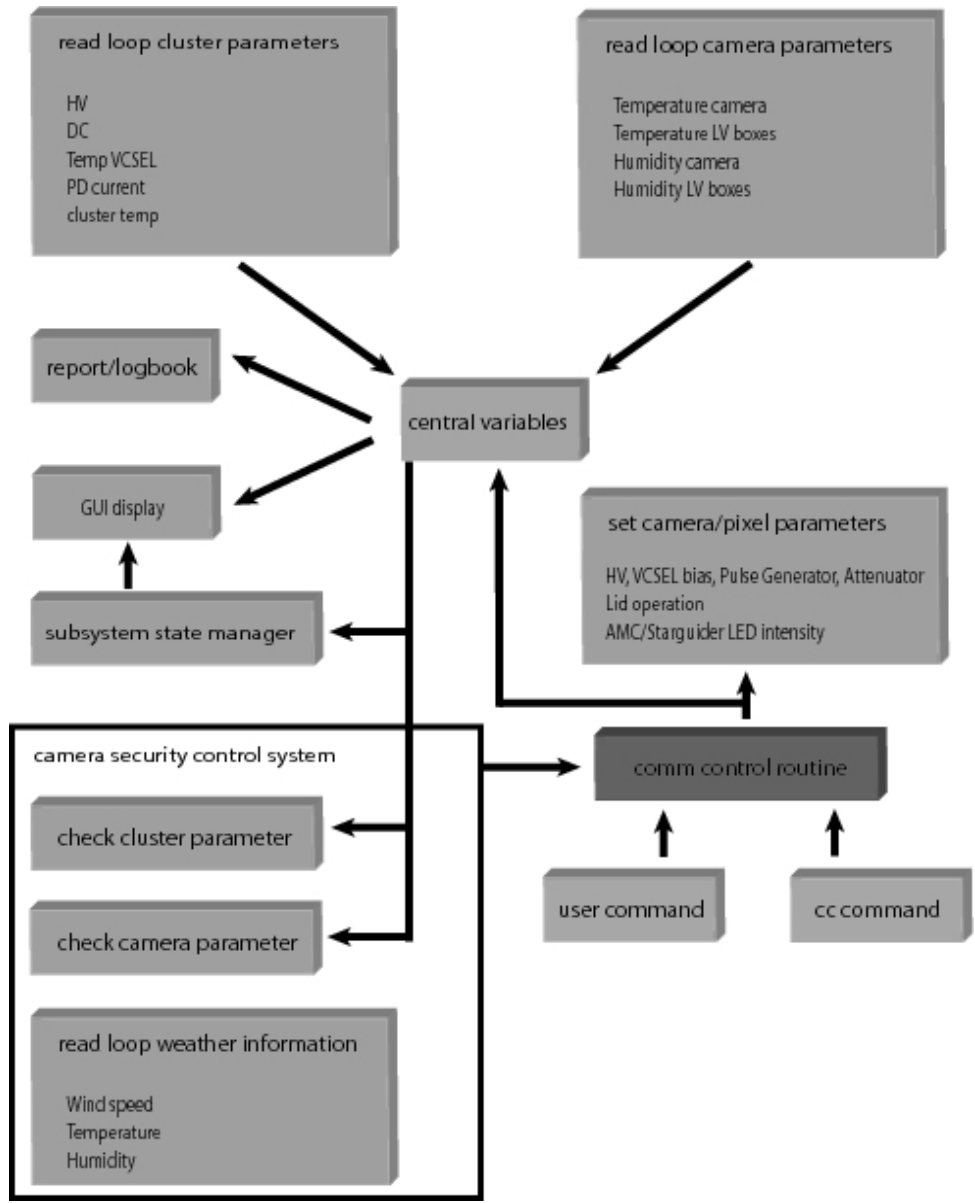}
  \caption{Scheme of the MAGIC-II camera control programm.}
  \label{caco_scheme}
 \end{figure*}

\section{Camera Control Program}
The \textit{Camera Control Program} is written in the visual programming language LabVIEW from National Instruments. This language provides a straightforward way to implement several control routines to a graphical and therefore user-friendly interface.\\
The main parts of the programm are the \textit{Central Variables File} and the \textit{Comm Control Routine} (cf. fig. \ref{caco_scheme}), which will be descibed in the following subsections. User commands can be given either via the GUI itself or via the supervising \textit{Central Control Program} (cf. sec. \ref{cc}). In addition, the Camera Control program disposes a security control routine, which checks permanently external and internal parameter and reacts, when necessary, on its own without any user interaction (cf. sec. \ref{cscs}).
\subsection{Central Variables File}
\label{cvf}
All available read-out information of each pixel, such as HV, DC current, VCSEL temperature and PD current, and general read-out information about the camera, such as camera temperature and humidity, temperature and humidity of the LV power supply boxes, is written to the \textit{Central Variables File} (cf. fig. \ref{caco_scheme}). In addition, the \textit{Central Variables File} also contains the desired values of the HV, VCSEL bias, attenuators and test pulse generators as given in the user settings together with the lid status and the AMC and Starguider LED intensity.\\
The \textit{Central Variables File} provides the GUI display, the report routine to the supervising Central Control Program (cf. sec. \ref{cc}) and the logbook with the relevant information. So, for example, the information shown in the GUI display is given from the \textit{Central Variables File}. Also the \textit{Camera Security Control System} (cf. sec. \ref{cscs}) is provided by the \textit{Central Variables File}.

\subsection{Comm Control Routine}
The \textit{Comm Control Routine} permanently checks the user input and the \textit{Camera Security Control System}. If any changes are given, the \textit{Comm Control Routine} sets the desired values, such as HV, VCSEL bias, pulse generator, attenuators, AMC or Starguider led intensity, or actions, such as lid operation or switching on/off the LV power supply (cf. fig. \ref{caco_scheme}). In addition, the \textit{Comm Control Routine} writes the new information to the \textit{Central Variables} file.

\subsection{Camera Security Control System}
\label{cscs}
The \textit{Camera Control Program} includes a safety routine, which shall protect the camera from misuse by accidential commands, from bad weather conditions and from hardware errors. This \textit{Security Control System} is permanently checking the pixel and camera parameter and the external weather information, such as wind speed, temperature and humidity. If a parameter exceeds the chosen limits, the \textit{Security Control System} reacts without any user interaction by sending the necessary commands to the \textit{Comm Control Routine} (cf. fig. \ref{caco_scheme}). E.g., an exceeding windspeed leeds to an automatically closing of the camera lids.

\subsection{Interaction with the Central Control Program}\label{cc}
The global control of the telescope is the supervising \textit{Central Control Program}. The \textit{Camera Control Program}, described in this paper, is an independent subsystem of the global control chain.\\
User commands, when given via the \textit{Central Control Program}, are sent to \textit{Camera Control} via ASCII text string. This string is processed within the \textit{Camera Control Program} and leads to the same actions as if the commands were given directly via the \textit{Camera Control} GUI interface.\\
\textit{Camera Control} permanently sends a report string to the \textit{Central Control Program} out of the \textit{Central Variables File} (cf. sec. \ref{cvf}). So the \textit{Central Control} gets all essential information about the camera and is able to monitor the camera in a sufficent way.

\subsection{Initializing Routine}
Starting the \textit{Camera Control Program} runs at first the \textit{Initializing Routine}. Here the important setup files, such as limits of parameters and masked clusters, are read and a first status check of the camera is processed. Anomalies, e.g. an opened lid, are reported to the user via pop-up windows. So the user is provided with the important information about the camera before starting operations. 
 
\section{Summary}
The \textit{Camera Control Program} is an independent subystem of the MAGIC-II control software chain. With its main parts, the \textit{Central Variable File} and the \textit{Comm Control Routine} it is an effective software to monitor and control and the various camera and pixel values and settings. Monitoring information is shown on the GUI display and reported to the supervising \textit{Central Control Program}. User input can be given either directly via the \textit{Camera Control} GUI interface or the \textit{Central Control Program}. A logbook records the camera information via ASCII text file. 
In addition, the \textit{Camera Control Program} includes a safety routine, which protects the camera from misuse by accidential commands, from bad weather conditions and from hardware errors.

\section*{Acknowledgments}
We would like to thank the Instituto de Astrofisica de Canarias for the excellent working conditions at the Observatorio del Roque de los Muchachos in La Palma. The support of the German BMBF and MPG, the Italian INFN and Spanish MICINN is gratefully acknowledged. This work was also supported by ETH Research Grant TH 34/043, by the Polish MNiSzW Grant N N203 390834 and by the YIP of the German Helmholtz Gemeinschaft.

\end{document}